\documentclass[floatfix,pre,twocolumn,showpacs,superscriptaddress]{revtex4}

\usepackage{acronym}
\usepackage{amsmath}
\usepackage{amssymb}
\usepackage{graphicx}
\usepackage{bm}

\def\dt{\delta_\mathrm{t}}
\def\gmc{\varGamma_\mathrm{g}}
\def\eNliq{e_N^\mathrm{L}}
\def\eNsol{e_N^\mathrm{S}}
\def\bcN{\beta_\mathrm{c}^N}

\begin{document}


\title{Separation and fractionation of order and disorder in
highly polydisperse systems}

\author{L.~A.~Fern\'andez} 
  \affiliation{Departamento de F\'\i{}sica Te\'orica I, Universidad
  Complutense, Av. Complutense, 28040 Madrid, Spain.}
  \affiliation{Instituto de Biocomputaci\'on y F\'{\i}sica de Sistemas
  Complejos (BIFI), Spain.}

\author{V.~Mart\'\i{}n-Mayor} 
  \affiliation{Departamento de F\'\i{}sica Te\'orica I, Universidad
  Complutense, Av. Complutense, 28040 Madrid, Spain.}
  \affiliation{Instituto de Biocomputaci\'on y F\'{\i}sica de Sistemas
  Complejos (BIFI), Spain.}

\author{B.~Seoane} 
  \affiliation{Departamento de F\'\i{}sica Te\'orica I, Universidad
  Complutense, Av. Complutense, 28040 Madrid, Spain.}
  \affiliation{Instituto de Biocomputaci\'on y F\'{\i}sica de Sistemas
  Complejos (BIFI), Spain.}

\author{P. Verrocchio} 
  \affiliation{Dipartimento di Fisica, Universit{\`a} di Trento, via Sommarive 14, 38050 Povo, Trento,
  Italy.}
  \affiliation{Instituto de Biocomputaci\'on y F\'{\i}sica de Sistemas
  Complejos (BIFI), Spain.}


\begin{abstract}
We study a polydisperse soft-spheres model for colloids by means of
microcanonical Monte Carlo simulations. We consider a polydispersity
as high as $24\%$. Although solidification occurs, neither a crystal
nor an amorphous state are thermodynamically stable. A finite size
scaling analysis reveals that in the thermodynamic limit: a) the
fluid-solid transition is rather a crystal-amorphous phase-separation,
b) such phase-separation is preceded by the dynamic glass transition,
and c) small and big particles arrange themselves in the two phases
according to a complex pattern not predicted by any {\em
  fractionation} scenario.
\end{abstract}


\maketitle


\acrodef{IS}{inherent structure}
\acrodef{RFOT}{random first-order theory}
\acrodef{CRR}{cooperatively rearranging regions}
\acrodef{MCT}{mode-coupling theory}
\acrodef{KCM}{kinetically constrained models}
\acrodef{FSS}{finite size scaling}
\acrodef{PSS}{polydisperse soft spheres}
\acrodef{FCC}{face centered cubic}
\acrodef{BCC}{body centered cubic}
\acrodef{MC}{Monte Carlo}
\acrodef{pdf}{probability distribution function}
\acrodef{PT}{Parallel Tempering}

\section{Introduction}

Although in condensed matter physics spatial order is naturally linked
to low temperatures, the presence of inherently disordered
interactions ({\em quenched} disorder) challenges such scenario. The
issue has been extensively addressed in lattice systems (spin glasses,
magnetic materials in random field, etc..) where quenched disorder in
fact inhibits spatially ordered structures (although not other types
of order). Much less is known about off-lattice systems. The issue
presents some practical consequences. For example crystallization of
very viscous colloidal samples with size dispersion $\delta$, see
Eq.~(\ref{DEF:DELTA}) below, larger than $12\%$ does not occur, even
after several months spent from the sample
preparation~\cite{poly:Pusey86}. This leads to several basic questions
about the equilibrium phase diagram of polydisperse
systems~\cite{poly:Bartlett98,poly:Kofke99,poly:Auer01,poly:Fasolo04,poly:Dullens04,poly:Chaudhuri05,poly:Fernandez07,colloids:Brambilla09,colloids:Zaccarelli09}. Does
enough large polydispersity hinder crystallization? Is the glass phase
stable rather than only metastable? Is there a dynamic interplay
between crystallization and the glass
transition~\cite{colloids:Brambilla09,colloids:Zaccarelli09}?  And,
probably at a more fundamental level, is thermodynamic equilibrium
relevant at all to describe real polydisperse materials or these are
instead inherently off-equilibrium over the experimental time scales?
Answering such questions is crucial for condensed matter physics,
since polydispersity is found both in artificial (synthetic colloids,
polymers) and natural systems, from supercooled liquids on the atomic
scale up to biological fluids such as blood.

An attempt to rationalize the experimental findings is the so-called
terminal polydispersity scenario where a characteristic value $\dt
\sim 0.12$ exists above which the homogeneous crystal becomes
thermodynamically unstable. There is not consensus however about what
kind of structure should replace such single phase crystal. Density
functional analysis~\cite{poly:Chaudhuri05} predicts the instability
of any crystal structure (even partial) above $\dt$, thus leaving the
amorphous ones (either liquid or solid) as the only possibility. Yet,
the moment free-energy approach~\cite{poly:Fasolo04} predicts {\em
  fractionation}: phase separation between many crystal phases [though
  of the same ordering, \ac{FCC}, for instance], each one with a much
narrower size dispersion than $\delta$. Fractionation is supported by
a recent numerical simulation that found that a first-order
fluid-solid transition actually occurs at any polydispersity
\cite{poly:Fernandez07}. However, the solid phase is quite complex, at
least in the high polydispersity region.  In fact, for $\delta > 0.19$
the transition regards only a fraction of the particles and the
ordered state is inhomogeneous. Such state has been previously
referred to as I-phase\cite{poly:Fernandez07}.

Here we study the high polydispersity region, in particular the point
$\delta=0.24$. The corresponding $\delta-\beta$ phase diagram ($\beta$ is the
inverse temperature, $1/T$) is sketched in the inset in
Fig.~\ref{SPINODAL}. This region is of great interest for various reasons.
First, the amount of crystalline order for the coldest/densest configurations
is unknown. It turns out to be phase-separated between a crystal and an
amorphous state. The pattern of particle-size distribution among the two
states does not follow any simple fractionation rule. Second, it has been
suggested \cite{poly:Fernandez07} that in this system the dynamic glass
transition occurs in the stable rather than in the metastable fluid
region. Our results support this claim in the large $N$ limit. Besides, the
detailed knowledge of the equilibrium structures is needed in order to design
new experimental or numerical methods to drive the system towards such
structures.

The layout of the rest of this work is as follows. In
Sect.~\ref{SECT:MODEL} we describe our model, the microcanonical
ensemble (Sect.~\ref{SECT:MICRO}), and the considered observables
(Sect.~\ref{SECT:OBSERVABLES}). Our simulation algorithm and our
thermalization checks are described in Sect.~\ref{SECT:NUM}. Our main
numerical results are described in Sect.~\ref{SECT:RESULTS}.  We
present our conclusions in Sect.~\ref{SECT:CONCLUSIONS}.

\section{Model}~\label{SECT:MODEL}
Take as a paradigm for polydisperse off-lattice systems the \ac{PSS}
model.  We consider particles of radius $\sigma_i\,,$ with
$i=1,2,\ldots,N\,$. The particle size $\sigma_i$ is drawn from a
\acf{pdf} $P(\sigma)$. Size polydispersity is in general characterized
by a single parameter, $\delta$, defined as the ratio among the
standard deviation and the mean of $P(\sigma)$:
\begin{equation}
\delta = \frac{\sqrt{\langle \sigma^2\rangle-\langle\sigma\rangle^2}}{\langle \sigma\rangle}\,.\label{DEF:DELTA}
\end{equation}
At least for small polydispersity, $\delta$ seems to be the only
feature of $P(\sigma)$ that controls the physical results.

Our particles interact via a pair potential:
\begin{eqnarray}\label{potential}
V_{ij}(r) &=& \epsilon \left(\frac{1}{x_{ij}^{12}} + x_{ij} -
\frac{13}{12^{12/13}}\right)\ \mathrm{if}\  x_{ij}<x_\mathrm{c}\,,\\
V_{ij}(r)&=&0 \ \mathrm{if}\  x_{ij}> x_\mathrm{c}\,,
\text{with\ } x_{ij}= \frac{r}{\sigma_i+\sigma_j} \,,\  
x_\mathrm{c}= 12^{\frac{1}{13}}.\nonumber
\end{eqnarray}
We take $\epsilon$ as energy unit. The potential is basically the repulsive
part of Lennard-Jones, $1/r^{12}$. The only role of the linear piece is to
provide a smooth long distance
cut-off~\cite{glass:Fernandez06c,algorithm:yan04}.

Our length unit, $\sigma_0$, is fixed by
\begin{equation}
\sigma_0^3 = \int \mathrm{d}\sigma_i
  \mathrm{d}\sigma_j P(\sigma_i) P(\sigma_j) (\sigma_i +
  \sigma_j)^3\ .
\end{equation}
Although ~(\ref{potential}) generalizes well known models for simple
liquids~\cite{hansen} (one would then choose $\sigma_0\sim 1$ nm), the scale
invariance of the $1/r^{12}$ potential suggests that our model may describe as
well colloids. For the colloidal case one would choose $\sigma_0\sim 1$
micrometer. In fact, the cutoff in the potential~(\ref{potential}) makes it
short-ranged as it is appropriate for colloidal systems.

We simulated $N$ particles in a box with periodic boundary conditions at
density $\rho=\sigma_0^{-3}$.  Due to the scale invariance of the $1/r^{12}$
potential, the thermodynamic parameter that controls the problem is the
combination $\varGamma \equiv \rho\: T^{-1/4}$ ($T$ is the
temperature)~\footnote{Strictly speaking, the long distance cut-off spoils
  scale invariance, so that one could question that $\varGamma$ is the
  controlling thermodynamic parameter. In practice, the cutoff is chosen to
  minimize its physical effects. In fact, the Mode Coupling
  transition~\cite{Goetze92} has been located with a variety of cut-off
  choices and polydispersities(see~\cite{Bernu87,Yu04,glass:Fernandez06} and
  present work). In all cases, when temperatures are expressed in terms of
  $\varGamma$, the location of the Mode Coupling transition
  $\gmc$ agreed to an accuracy of at least $1\%$.}.

Here we study the case where the size distribution is flat (constant
in the range
$[\sigma_\mathrm{min},\sigma_\mathrm{max}]$). Sample-to-sample
fluctuations are eliminated by picking the diameters in a deterministic way~\cite{poly:Santen01,
  poly:Fernandez07},
\begin{equation}
\sigma_i =
  \sigma_\mathrm{min} + (i-1)\frac{\sigma_\mathrm{max}-\sigma_\mathrm{min}}{N-1}\,.
\end{equation}
Observe that
\begin{equation}
\delta=\frac{1}{\sqrt{3}}\frac{(r-1)}{(r+1)},\ \mathrm{with}\ r=\frac{\sigma_\mathrm{max}}{\sigma_\mathrm{min}}\,.
\end{equation}
Hence, $\sigma_\mathrm{max}/\sigma_\mathrm{min} \to \infty$ at
$\delta_{\infty}=1/\sqrt{3}\approx 0.57735$.

Since polydispersity hampers crystallization~\cite{poly:Pusey86}, a glass
transition is to be expected. Although most of this work has been performed in
the microcanonical ensemble, let us mention that we have also estimated the
glass temperature in the $(N,V,T)$ ensemble by means of \ac{MC}
simulations. We simulated the equilibrium fluid state using only standard
Metropolis single-particle moves (different choices of microscopic dynamics
lead to basically equivalent results, see~\cite{berthier07}).  To locate the
kinetic glass transition by computing the relaxation time $\tau$ of the fluid
for $N=500,864$ in the range $\varGamma \in [1.3, \, 1.46]$ (data not
shown). Our definition of the kinetic glass transition $\gmc$ corresponds to
the point when $\tau$ surpasses the $10^6$ \ac{MC} steps. Both for $N\!=\!500$
and 864 particles, we find that $\gmc= 1.455(5)$.

The signification of $\gmc$ is rather different, depending on whether one is
studying liquids (i.e. $\sigma_0\sim$ 1 nm) or colloids ($\sigma_0\sim$ 1
micrometer). In the colloidal case, a standard \ac{MC} step corresponds
roughly to 0.01 seconds of experimental time~\cite{poly:Simeonova04}, so that
$\tau\sim 10^6$ \ac{MC} steps $\sim$ 3 hours of physical time and $\gmc$
corresponds to the experimental glass transition. On the other hand, for
liquids 1 \ac{MC} step is roughly equivalent to one picosecond. Thus,
$\tau\sim 10^6$ \ac{MC} steps $\sim 10^{-6}$ physical seconds, implying that
$\gmc$ rather corresponds to the Mode Coupling
transition~\cite{Goetze92}. Indeed, for most molecular and polymeric
glass-forming liquids $\tau$ at the Mode Coupling temperature lies in the
range $10^{-7.5}$ and $10^{-6.5}$ seconds~\cite{Novikov03}.

\subsection{The constant energy ensemble}\label{SECT:MICRO}

We shall be working in the $(N,V,E)$ ensemble. Specifically, we shall
be using Lustig's microcanonical Monte
Carlo~\cite{algorithm:lustig98} in the formulation
of~\cite{algorithm:martin-mayor07}. 

Let $U$ be the total potential energy of our system,
\begin{equation}
U=\sum_{i<j} V(r_{i,j})\,,\ (u\equiv U/N)\,. 
\end{equation}
Thus, the total energy is 
\begin{equation}
E=U+K\,,\ (e\equiv E/N)\,.
\end{equation}
where $K=\sum_{i=1}^N p_i^2/2$ is the kinetic energy associated to the conjugated
momenta $\{p_j\}$. Here, we are considering just one conjugated momentum per
particle. As the kinetic energy is non-negative by definition, we
should have $E\geq U$. The conjugated momenta are explictly integrated
out (they are simply a conceptual device to introduce the
ensemble~\cite{algorithm:lustig98}). 

A quantity of major importance in the microcanonical ensemble is the
entropy density, $s_N(e)$:
\begin{eqnarray}
\mathrm{exp}[N s_N(e)]&=& \frac{(2\pi N)^{N/2}}{N \Gamma(N/2)}\times\\  
&\times&\int \frac{\prod_{i=1}^N \,\mathrm{d}{\boldsymbol r}_i }{N!}  (e-u)^{\frac{N}{2}
-1} \theta(e-u)\,.\nonumber
\end{eqnarray}
The Heaviside step function, $\theta(e-u)$, enforces $e>u$. The
microcanonical average of an arbitrary function of the particle
positions $\{ {\boldsymbol r}\}_i$ and of the energy density $e$,
$O(\{ {\boldsymbol r}\}_i;e)$ is
defined as 
\begin{eqnarray}
\langle O\rangle_e&\equiv&   \frac{\int \prod_{i=1}^N \,\mathrm{d}{\boldsymbol r}_i \, O(\{ {\boldsymbol r}\}_i;e)  \omega_N (\{ {\boldsymbol r}\}_i;e)}
{\int \prod_{i=1}^N \,\mathrm{d}{\boldsymbol r}_i \, \omega_N (\{ {\boldsymbol r}\}_i;e)}\,,
\end{eqnarray}
where,
\begin{eqnarray}
\omega_N (\{ {\boldsymbol r}\}_i;e)&=& (e-u)^{\frac{N}{2}
-1} \theta(e-u)\,.\label{DEF:MICROWEIGHT}
\end{eqnarray}

\subsection{Observables}\label{SECT:OBSERVABLES}

\subsubsection{The inverse temperature}
The main observable in a microcanonical simulation is the inverse
temperature, computed as a microcanonical expectation value at fixed
energy $e$:
\begin{equation}
\beta(e)\equiv \langle\hat\beta\rangle_e,\quad \hat\beta
=\frac{N-2}{2N (e-u)}\,.
\label{BETADEF}
\end{equation}
Note that
\begin{equation}
\beta(e) = \frac{\mathrm{d} s_N(e)}{\mathrm{d} e}\,.
\end{equation}

The function $\beta(e)$ holds the key to connect the microcanonical
formalism with the canonical one. Indeed, the {\em canonical}
probability density for $e$,
$P_{\beta}^{(N)}(e)\propto\mathrm{exp}[N(s_N(e)-\beta e)]$ can be recovered
from $\beta(e)$:
\begin{equation}
\log P_{\beta}^{(N)}(e_2)-\log P_{\beta}^{(N)}(e_1)=
N\int_{e_1}^{e_2}\mathrm{d}e\, \left(
\beta(e) -\beta\right)\,.\label{LINK}
\end{equation}
In the {\em thermodynamically stable region} (i.e.  $\mathrm{d}
\beta(e)/\mathrm{d}e <0$), there is a single root of $\beta(e)=\beta$,
located at the value of $e$ where $P_{\beta}^{(N)}(e)$ is maximum. Instead, at
phase coexistence there are several solutions for
$\beta(e)=\beta$. Their interpretation is explained in
Sect.~\ref{SECT:MAXWELL}. 

\subsubsection{The particle-density field}

As we mentioned in the Introduction, we expect large particle-density
fluctuations. In order to detect them, we study the
Fourier-transformed density field at the smallest, non-vanishing
wavenumber allowed by the periodic boundary conditions:
\begin{equation}
{\cal F} \equiv \frac{1}{3}\left(|\hat\rho (2\pi/L,0,0)|^2+\text{
  permutations}\right)\,,
\label{eq:F}
\end{equation}
where $L$ is the linear dimension of our cubic
simulation box and the Fourier field is
\begin{equation}
\hat\rho(\bm q)=\frac{1}{N}\sum_{i} e^{{\mathrm i}\bm q\cdot \bm r_i}\,.
\end{equation}
Note that $\hat\rho(\bm q)$, a function of the particles configuration, yields
the static structure factor through $S(\bm q)=N\langle|\hat\rho(\bm
q)|^2\rangle\,$.  In particular, $\hat\rho(0)$ is our non-fluctuating particle
density $\rho$.

\subsubsection{Crystalline order parameters}

In order to study simultaneously crystallization and fractionation, we
generalize the (rotationally-invariant) crystal order
parameters\cite{Steinhardt83,Wolde96} by measuring the crystal order
only of a given set of particles ${\cal I}(x)$ (namely, particles
whose index $i$ verifies $|i-xN|<0.05N$, hence only particles of
similar size are considered):
\begin{equation} 
Q_l(x) \equiv \left( \frac{4 \pi}{2l +1} \sum_{m = -l}^{l} \left|
Q_{lm}(x) \right|^2 \right)^{1/2},
\label{Qs}
\end{equation}
where ($Y_{lm}$ are the spherical harmonics):
\begin{equation}
Q_{lm}(x) \equiv 
\frac{\sum_{\sigma_i \in {\cal I}(x)}\, q_{lm}(i)}{\sum_{\sigma_i \in {\cal I}(x)}
N_b(i)},\, q_{lm}(i) \equiv \sum_{j=1}^{N_b(i)} Y_{lm}({\hat r_{ij}}).
\end{equation}
The index $j$ in the latter sum runs over the $N_b(i)$ neighbors of
the particle $i$ and $\hat r_{ij}$ is the unit vector linking the
position of particles $i$ and $j$. Particles $i$ and $j$ are said to
be neighbors if $||\bm r_i - \bm r_j||<\varDelta$. In order to
meaningfully fix the scale $\varDelta$, we considered the average
number of neighbors as a function of $\varDelta$, finding a
plateau. The height of the plateau is remarkably $N$-independent, but
its {\em width} increases with $N$ (so, the particular choice of
$\varDelta$ becomes less critical as $N$ grows). We fixed the value
$\varDelta=0.35$ (in units of the maximum cut-off for the potential
  $2\sigma_\mathrm{max}\, x_\mathrm{cut}$), that lies in the plateau for all our
values of $N$ and for all our energies in the solid phase.

Since we let the
fraction of particles be a finite fraction $x$ of $N$ the $Q_l$'s
are intensive quantities. In amorphous phases $Q_l(x)$ decrease as
$1/\sqrt{N}$ while in crystalline ones they remain of order $1$. In
particular, we consider the case $l=6$.

\section{Numerical Algorithms and thermalization tests}\label{SECT:NUM}

In order to study the fluid-solid phase transition we implement a
microcanonical \ac{MC}
strategy\cite{algorithm:martin-mayor07,algorithm:lustig98}.  Fixing
the total energy density $e$, while the temperature and the potential
energy fluctuate (see Eq.\eqref{BETADEF}), we follow the evolution
from one phase to the other by studying $e$ in the {\em energy gap}
between the two phases.  This strategy turned out to be essential to
assess the first-order nature of the phase transition in disordered
Potts models\cite{Potts:Fernandez08}.

The peculiarity of the polydisperse models addressed here, as compared
with Potts and similar models, is in that the phase transition
actually corresponds to a phase separation.  In fact, our low energy
state is inhomogeneous\cite{poly:Fernandez07}. Thus moving $e$ from
large values (fluid) to small ones (partly solid) we gently {\em
  accompany} the system during the growth of the spatially segregated
regions. Although internal energy will not be the only reaction
coordinate (see below), we have found useful to combine microcanonical
\ac{MC} with a modified \acf{PT} algorithm~\cite{hukushima:96,marinari:98b}.

For the sake of clarity, we divide the remaining part of this Section
in three paragraphs: particle movements at fixed energy
(Sect. \ref{SECT:FIXED-E}), Parallel Tempering (Sect.~\ref{SECT:PT}),
and thermalization checks (Sect.~\ref{SECT:THERMALIZATION}).

\subsection{Particle movements at fixed energy}\label{SECT:FIXED-E}

The particle moves at fixed energy were, with $50\%$ probability,
either standard Metropolis single-particle moves, or global swap
attempts (modified for a polydisperse system). Let us recall that in a
swap move, one attempts to exchange the position of two particles of
different sizes~\cite{algorithm:grigera01}. Both for single-particle
and for swap moves we compute the ratio of the microcanonical weights,
defined in Eq.(\ref{DEF:MICROWEIGHT}), for the new and the old
configuration $\omega_N^\mathrm{old}/\omega_N^\mathrm{new}$. The new
configuration is accepted with Metropolis probability
$\mathrm{min}\{1,\omega_N^\mathrm{old}/\omega_N^\mathrm{new}\}$.

To fully describe the swap algorithm, we need to discuss how we choose
the pair of particles, $A$ and $B$, whose position we are trying to
interchange. Note that one needs to balance two effects in
polydisperse systems.  The acceptance is larger the closer the two
particle sizes are. However, exchanging very different particles
produces a more significant effect when trying to equilibrate the
system. Our compromise has been the following. We pick particle $A$
with uniform probability over the $N$ possibilities.  We pick $B$ with
uniform probability among particles such that $|\sigma_B
-\sigma_A|<0.2 (\sigma_\text{max}-\sigma_\text{min})$ . Particle $B$
is accepted with probability 1 if $|\sigma_B -\sigma_A|> 0.1
(\sigma_\text{max}-\sigma_\text{min})$ or with probability 0.2 in the
opposite case. In case of rejection, a new particle $B$ is selected
until a suitable candidate is picked.

On the coexistence-line, swap moves reduce by three orders of
magnitude the tunneling time between the fluid and the solid
phase.

\subsection{The microcanonical parallel tempering}\label{SECT:PT}

In our Parallel Tempering simulations, several statistically
independent copies of the system at different energies are simulated
(fixed {\em energies} rather than fixed temperatures, as it is
normally performed in standard
\ac{PT}~\cite{hukushima:96,marinari:98b}).

Each Monte Carlo time unit consists of two steps:
\begin{enumerate}
\item For each copy of the system, we perform $10^5\times N$ particle
  move attempts at fixed energy (either single-particle displacements
  or particle-swap attempts). During this stage, each copy of the
  system is completely independent from the others.

\item Copies of the system at neighboring energies try to exchange
  their particle configuration. We first try to sweep the two
  configurations at the lowest energy, afterwards the second lowest
  with third lowest, etc.  In this way, the particle-configuration at
  the lowest energy has a chance of getting to the highest energy in a
  single sweep.

For the sake of clarity let us name $A,B$ the two systems that are
currently attempting to exchange their particle configuration. The
exchange is accepted with probability
\begin{equation}
\mathrm{min}\left[1\ ,\ \frac{\omega_N(\{{\boldsymbol r}_i^{(A)}\};e^{(B)})\,\omega_N(\{{\boldsymbol r}_i^{(B)}\};e^{(A)})}
{\omega_N(\{{\boldsymbol r}_i^{(A)}\};e^{(A)})\, \omega_N(\{{\boldsymbol r}_i^{(B)}\};e^{(B)})}\right]\,.
\end{equation}
The microcanonical weights $\omega_N$ are given in
Eq.(\ref{DEF:MICROWEIGHT}).
\end{enumerate}
Further details on the simulation are summarized in Table~\ref{tabla}.

Let us finally note that the here used Monte Carlo method is quite similar to
that of Refs.~\cite{algorithm:yan03,algorithm:yan04}. We briefly mention the
main differences. First, particle swap at fixed energy was not used in
Refs.~\cite{algorithm:yan03,algorithm:yan04}. Second, phase coexistence (and
the related Maxwell construction) was not studied. Third, in the formulation
of~\cite{algorithm:yan03}, one has a single copy of the system that performs a
random-walk in energy space: it is a sort of simulated annealing
simulation~\cite{marinari:98b}, rather than our parallel tempering. Besides,
the approximation $\beta(e)\approx (N-2)/[2N \langle (e-u)\rangle]$ is used,
which coincides with Eq.~(\ref{BETADEF}) only up to corrections of order
$1/N$. The formulation of~\cite{algorithm:yan04} is somehow intermediate
between simulated annealing and parallel tempering. The energy range of
interest is spliced into non-overlapping subranges. Each copy of the system is
assigned to an energy subrange, where it performs a simulated annealing. From
time to time one uses parallel tempering to exchange the copies of the system
attached to neighboring energy subranges.

\subsection{Thermalization checks}\label{SECT:THERMALIZATION}

\begin{figure}[h]
\includegraphics[angle=270,width=\columnwidth,trim=28 40 21
  40]{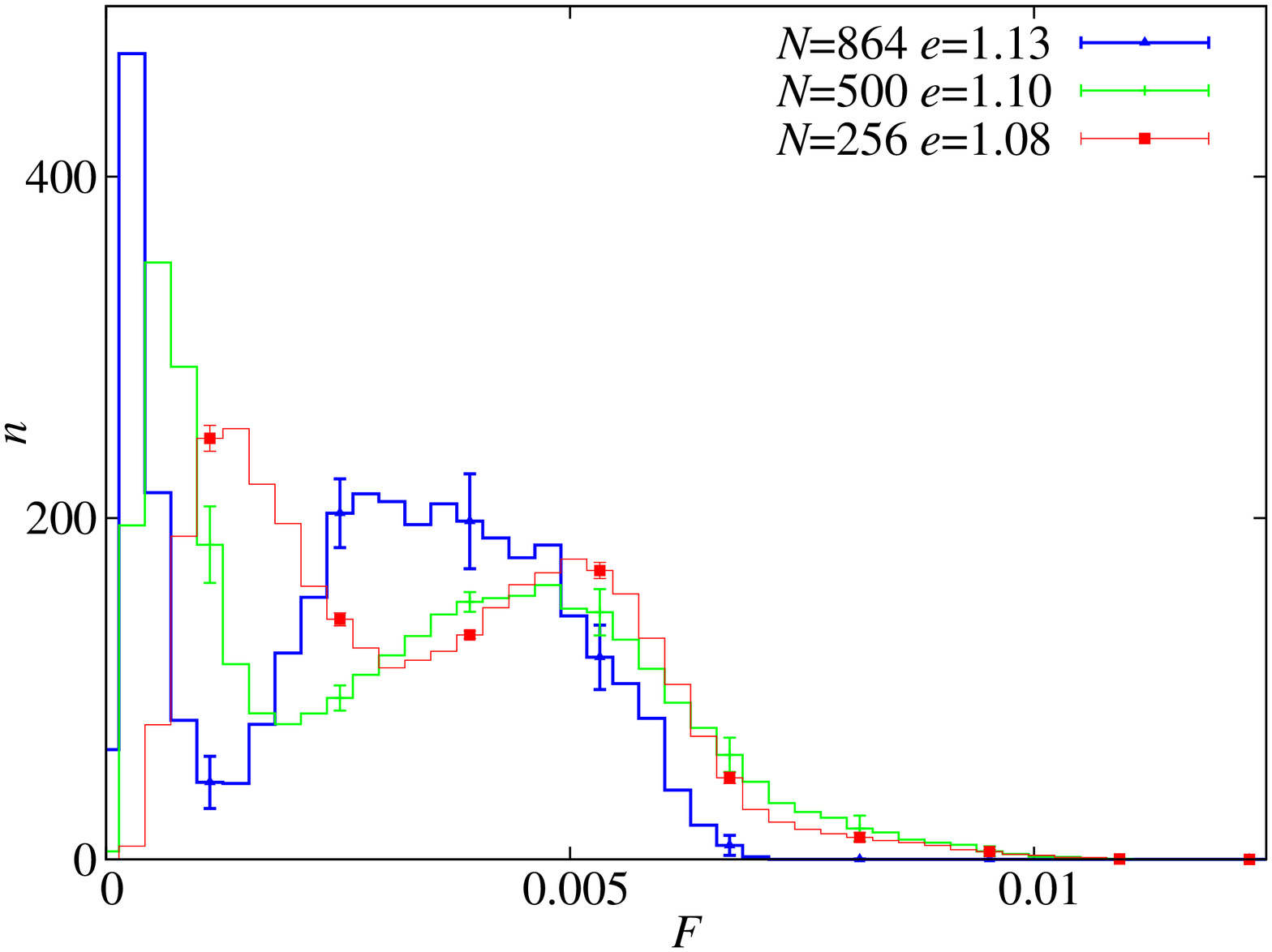}
\includegraphics[angle=270,width=\columnwidth,trim=8 40 21
  40]{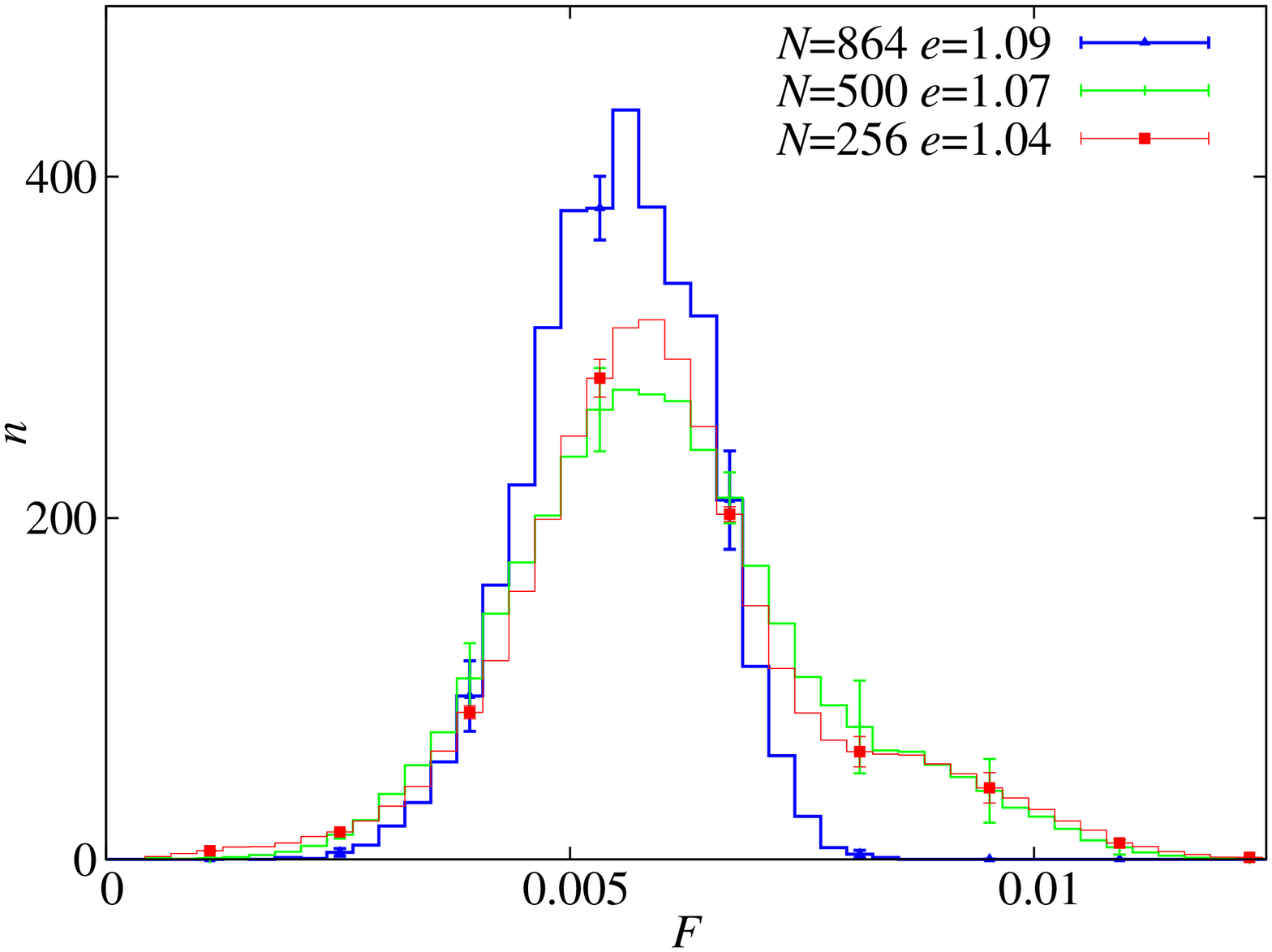}
\caption{\ac{pdf} of ${\cal F}$, Eq.\eqref{eq:F} at
  various representative values of $e$. Data in the top panel are
  computed at energy densities in the energy gap between the fluid and
  the solid phases. The double peak structure reveals phase
  coexistence (the position of the leftmost peak scales as
  $1/N$). Data in the bottom panel are computed for $e$ in the solid
  phase (the $e$-dependency there is very mild).}
\label{histogramas}
\end{figure}

\begin{figure}
\includegraphics[angle=270,width=\columnwidth,trim=28 40 21 40]{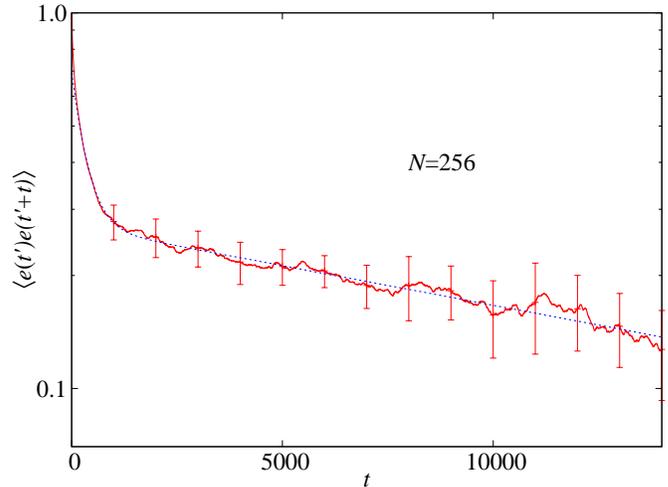} 
\caption{The
  (connected) time autocorrelation function for the energy in the \ac{PT}
  can be fitted (dotted line) as $\langle e(t') \, e (t' +t)\rangle= a_{FS}\mathrm{e}^{-t/\tau_{FS}}+ a_{SS} \mathrm{e}^{-t/\tau_{SS}}\,.$}
\label{correlaciones}
\end{figure}

A crucial issue of \ac{PT} simulations is to ensure thermalization.
Fortunately, a nice feature of \ac{PT} is that it provides a sound
thermalization check by controlling that all systems visit uniformly
the whole range of energies\cite{PTcheck:fernandez09}.

At variance with the Potts case, phase coexistence {\em inside the
  energy gap} between a fluid and a solid is apparent from the
\ac{pdf} of the quantity ${\cal F}$, defined in Eq.~\eqref{eq:F}. Our
results are shown in Fig.~\ref{histogramas}-top). At values of $e$
close to the transition, we identify two coexisting peaks. One of them
is located at ${\cal F}\sim 1/N$, as expected for an homogenous fluid
phase. On the other hand, the position of the large ${\cal F}$ maximum
becomes $N$-independent (this is clearer at lower energies, see bottom
panel in Fig.~\ref{histogramas}), as it should occur for an
inhomogeneous solid.  Such phase coexistence makes us to expect a
large growth with $N$ of the autocorrelation
times\cite{LandauBinder}. Actually, the \ac{pdf} for ${\cal F}$ at low
energies (Fig.~\ref{histogramas} bottom) displays a shoulder at large
${\cal F}$, which corresponds to even more inhomogeneous
solids. Hence, the \ac{PT} dynamics is ruled by two different
processes: tunneling from fluid to solid, and a second tunneling to
even more inhomogeneous configurations.

The random-walk in the energy space is best described through a \ac{PT}
time autocorrelation function (see Ref.~\cite{PTcheck:fernandez09} for
details), that indeed can be fit to a double exponential for $N=256$
and $N=500$, see Fig.~\ref{correlaciones}. Mind that the {\em time} in
this correlation functions correspond to the time-unit defined in
Sect.~\ref{SECT:PT}. It is not related to any physical
time-correlation.

As expected from the above discussion, we identify two different time scales
in Table~\ref{tabla}, one associated to the coexistence of the homogeneous and
inhomogeneous phase, $\tau_{FS}$, and a larger time, $\tau_{SS}$, related to
the more inhomogeneous configurations. For $N=864$, we could only identify the
$\tau_{FS}$ scale. Probably, $\tau_{SS}$ is larger than the total time in our
simulation. We remark that $\tau_{FS}$ for $N\!=\!256$ can be estimated with a
$5\%$ accuracy, while only the order of magnitude of $\tau_{SS}$ is
determined. We have explicitly checked that the effects of these very
inhomogeneous configurations on the Maxwell construction is fortunately
smaller than our statistical errors~\footnote{This is doing by following the
  random-walk of each copy of the system in energy space. One easily realizes
  that, along the simulation, the system switches between {\em trapped} and
  ergodic phases.  During a {\em trapped} phase, one or more copies of the
  system remains confined at the lowest energies and displays larger values of
  ${\cal F}$. In fact, the characteristic time $\tau_{SS}$ corresponds to the
  average duration of the trapped phase. The statistical analysis can be done
  either considering the full simulation or only the ergodic pieces of it. The
  Maxwell construction comes out compatible within statistical errors.}.
Furthermore, from the point of view of our measured crystalline order
parameters (see below), the more inhomogeneous configurations are not
distinguishable from the main peak in the \ac{pdf}.

\begin{figure}
\includegraphics[angle=270,width=\columnwidth,trim=27 25 21 33]{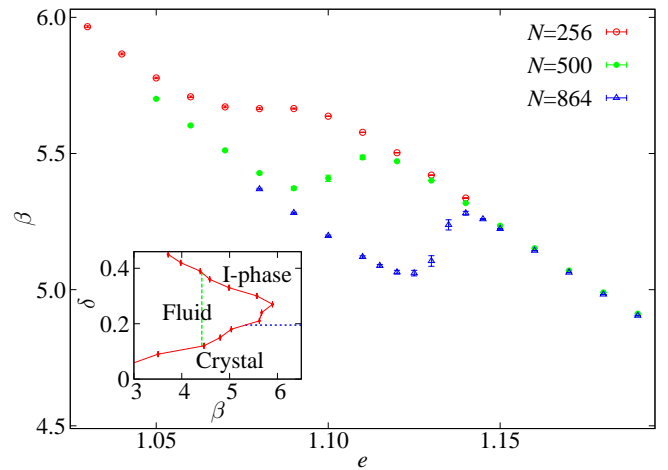} 
\caption{Finite size effects in the Maxwell construction. Main panel:
  the inverse temperature $\beta(e)$ as a function of the energy
  density $e$ for various sizes of the sample. Inset: $\delta-\beta$
  phase diagram of the system obtained from the data in
  Ref.~\onlinecite{poly:Fernandez07}.}
\label{SPINODAL}
\end{figure}
\section{Numerical Results}\label{SECT:RESULTS}

\subsection{The Maxwell construction}\label{SECT:MAXWELL}
As was mentioned in Sec.~\ref{SECT:OBSERVABLES}, in a microcanonical simulation, a quantity of major interest is the
(inverse) temperature, $\beta(e)$, see
Eq.~(\ref{BETADEF}). Thermodynamic stability requires that $\beta(e)$
be a decreasing function (i.e. positivity of the specific heat). Yet,
see main panel in Fig.~\ref{SPINODAL}, this is not the case close to a
first-order phase transition. The lack of monotonicity can be used to
obtain the critical temperature, surface tension, etc. through Maxwell
construction (see below, and Ref.\cite{algorithm:martin-mayor07} for
details). Generally speaking, $\beta(e)$ has two distinct branches,
one describing the fluid and the other the solid phase, where the
specific heat $C_v \equiv - \beta^2 d e /d \beta$ is positive,
connected by a thermodynamically instable line where $C_v
<0$. Although at finite $N$ the system does not undergo a real phase
transition, there are various criteria to define an (inverse) critical
temperature, $\bcN$, where the two different phases coexist with the
same thermodynamic weight. Here we utilize the Maxwell construction,
which amounts to obtain $\bcN$ as a solution of:
\begin{equation}
0=\int_{\eNsol(\bcN)}^{\eNliq(\bcN)}
\mathrm{d}e\, \left( \beta(e) -\bcN\right)\,, \label{MAXWELL}
\end{equation}
where the energy $\eNliq(\bcN)$ ($\eNsol(\bcN)$) in turn corresponds
to the rightmost (leftmost) root of the equation $\beta(e)=\bcN$.
Eq.(\ref{LINK}) shows that the Maxwell constructions amounts to the
famous equal-height rule for the canonical probability-distribution
function $P_\beta(e)$.

In Fig.~\ref{SPINODAL} we show the function $\beta(e)$ for $N=256,500,864$.
At odds with other models displaying a first order transition, as $N$ grows,
both the supercooled fluid (fluid branch with $\beta > \bcN$) and the
overheated solid (solid branch with $\beta < \bcN$) lines become longer.

As for the values of $\bcN$ reported in Table~\ref{tabla}, they decrease with
$N$. Asymptotically, finite $N$ corrections are of order $1/N$ (see
\cite{algorithm:martin-mayor07} and references therein). A fit
$\bcN=\beta_\mathrm{c}^{\infty}+ a_1/N$ fails badly the $\chi^2$ test. In
other words, our estimates for $\bcN$ are accurate enough to resolve
subleading scaling corrections in $1/N$. Thus, we have used a different
approach. Let us assume that scaling corrections take the form of a smooth
function in $1/N$, $\bcN=\beta_\mathrm{c}^{\infty}+ a_1/N+a_2/N^2+\ldots$. If
we have at our disposal three values of $N$, we may compute a quadratic
estimator (exact, up to corrections of order $1/N^3$):
\begin{eqnarray}
\beta_\mathrm{c}^{\infty,\mathrm{quad}}&=&\beta_\mathrm{c}^{N_1}\frac{N_1^2}
{(N_1-N_2)(N_1-N_3)}+\nonumber\\ 
&+&\beta_\mathrm{c}^{N_2}\frac{N_2^2}
{(N_2-N_1)(N_2-N_3)}+\nonumber\\ 
&+&\beta_\mathrm{c}^{N_3}\frac{N_3^2}
{(N_3-N_1)(N_3-N_2)}\,.
\end{eqnarray}
Computing the {\em statistical} error in
$\beta_\mathrm{c}^{\infty,\mathrm{quad}}$ is trivial, since
$\beta_\mathrm{c}^{N_1}$, $\beta_\mathrm{c}^{N_2}$ and
$\beta_\mathrm{c}^{N_3}$
are statistically independent random variables. Using the data in
Table~\ref{tabla} we get
\begin{equation}
\beta_\mathrm{c}^{\infty,\mathrm{quad}}= 4.624(20)\,,\quad
\varGamma_\mathrm{c}^{\infty,\mathrm{quad}}=1.4664(15)\,.
\end{equation}
However, the quadratic polynomial in $1/N$ that interpolates our values
$\beta_\mathrm{c}^{N_1}$, $\beta_\mathrm{c}^{N_2}$ and
$\beta_\mathrm{c}^{N_3}$ displays a maximum by $N\approx 256$, and decreases
for smaller $N$. Hence, $\beta_\mathrm{c}^{\infty,\mathrm{quad}}$ probably
overemphasizes curvature effects. On the other hand, a  linear (in $1/N$) extrapolation from $N_1=864$
and $N_2=500$ yields
\begin{equation}
\beta_\mathrm{c}^{\infty,\mathrm{linear}}= 4.791(11)\,,\quad
\varGamma_\mathrm{c}^{\infty,\mathrm{linear}}=1.4795(9)\,.
\end{equation}
The correct thermodynamic limit probably lies in between of the two estimators
$\varGamma_\mathrm{c}^{\infty,\mathrm{quad}}$ and
$\varGamma_\mathrm{c}^{\infty,\mathrm{linear}}$, above the kinetic glass
transition at $\gmc=1.455(5)$.

Furthermore, $\beta(e)$ also allows us to compute the surface
tension\cite{algorithm:martin-mayor07},
\begin{equation}
  \bcN\sigma_0^2\Sigma^{(N)}=\frac{N}{2L^{D-1}}
    \int_{e^*_N(\bcN)}^{\eNliq(\bcN)}
    \mathrm{d}e\, \left( \beta(e) -\bcN\right)\,,\label{SIGMAEQ}
\end{equation}
[recall that equation $\beta(e)=\bcN$ has three solutions
$\eNsol(\bcN)<e^*_N(\bcN)<\eNliq(\bcN)$].  Data is shown in Table~\ref{tabla}.
\begin{table*}
\centering
\begin{tabular*}{\columnwidth}{@{\extracolsep{\fill}}ccccccc}
\hline
$N$ & $\beta_\mathrm{c}$ & $\varGamma_\mathrm{c}$ & $\Sigma^{(N)}\bcN\sigma_0^2$&$\tau_{FS}$&$\tau_{SS}$&$L_\mathrm{sim}$ \\
\hline
$256$ & 5.665(3) &  1.5428(2)  &   --- &317(15) &$\sim$20000&$5\!\times 32000$\\
$500$ & 5.432(5)  &  1.5267(3)  & 0.0035(2)&$\sim$1000& $\sim$15000&$2\!\times\!30000$     \\
$864$ & 5.162(4)  &  1.5073(3)  & 0.0088(4)&$\sim$7000&---&$1\!\times\!12000$    \\
\hline
\end{tabular*}
\caption{Parameters of simulations and Maxwell construction.  For each
  number of particles, $N$, we estimate two characteristic time scales
  $\tau_{FS}$ and $\tau_{SS}$ for the \ac{PT} random walk in energy
  (see text), in units of \ac{PT} attempts.  We perform $10^5 N$
  \ac{MC} steps at fixed energy, then try a \ac{PT} sweep. We also
  report the total length of our simulations in units of \ac{PT}
  sweeps ($5\times 32000$ stands for 5 independent runs of $32000$
  \ac{PT} sweeps each).  The energies chosen for the \ac{PT} were
  evenly spaced $e_{i+1}-e_i=0.01$, in the intervals $[0.95,1.14]$
  ($N=256$), $[1.05,1.2]$ ($N=500$) and $[1.08,1.19]$ ($N=864$). For
  $N=864$ we added to the \ac{PT} energy list the values
  $1.115,1.125,1.135$ and $1.145$ in the fluid-solid energy gap. We
  also report the (inverse) critical temperature (and the associated
  $\varGamma_\text{c}=\rho \beta_\text{c}^{1/4}$), as well as the
  dimensionless surface tension $\Sigma\bcN\sigma_0^2$, as computed
  from Maxwell's construction.}
\label{tabla}
\end{table*}

\begin{figure*}
\includegraphics[angle=270,width=0.7\textwidth]{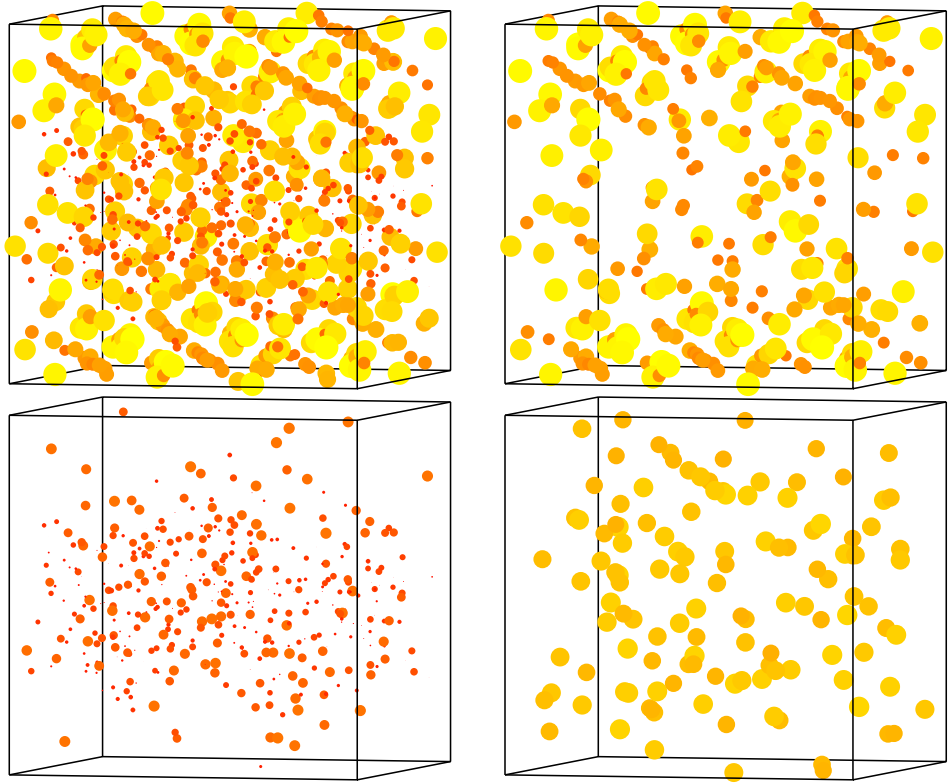} 
\caption{Snapshot of a typical low energy configuration ($N\!=\!864$,
  $e\!=\!1.01$). Top-left: whole system. Top-right: particles with index
  $i\!>\!725$ and $i \in \left[400,600\right]$. Bottom-left: particles
  $i\!<\!400$. Bottom-right: particles $i \in \left[600,725\right]$. The size
  of the circles are proportional to the particle sizes.}
\protect{\label{PICTURES}}
\end{figure*}

\subsection{Fractionation and crystalline ordering}

The need for generalized order parameters, Eq.~\eqref{Qs}, follows
from visual inspection of a typical $N=864$ low-energy configuration,
see Fig.~\ref{PICTURES}. In fact, the smallest $400$ particles
(particle index $i <400$) and some of the intermediates ($i \in \left[
  600, 725\right]$) show no sign of spatial order (bottom), while
particles with $i>725$ and $i \in \left[ 400, 600\right]$ form
crystalline planes. Ordered and disordered particles fill different
regions of the sample.

\begin{figure}
\includegraphics[angle=270,width=\columnwidth,trim=28 25 21 29]{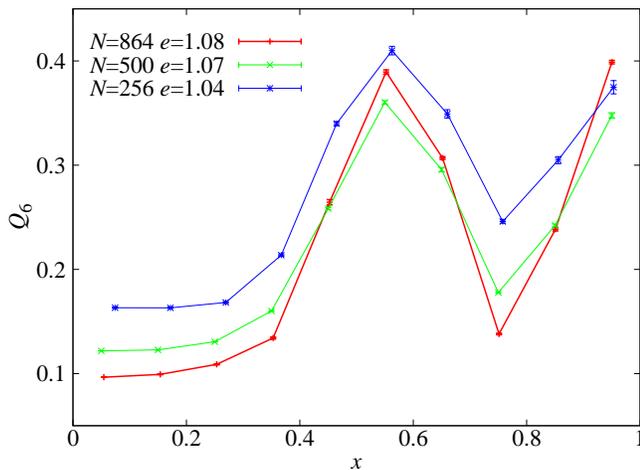} 
\caption{The crystal order parameter $Q_6(x)$, Eq.\eqref{Qs} as a
  function of the particles size $x$, for different $N$ values.}
\label{FRACTIONATION}
\end{figure}

Our results in Fig.~\ref{FRACTIONATION} confirm this picture. For
$x<0.45$ the crystalline order parameters decay as $1/\sqrt{N}$, while
for $x\!=\!0.55$ and $x\!=\!0.95$ we obtain results roughly $N$
independent. Thus, while the latter group of particles form a crystal
($Q_6$ is somewhat smaller than expected for \ac{FCC} ordering), the
former one remains amorphous. As for polydispersities, in the
two-components crystal we estimate that $\delta\sim0.15$, while in the
fluid $\delta\sim 0.24$.

\section{Conclusions}\label{SECT:CONCLUSIONS}

In summary, we have studied in the microcanonical ensemble a
soft-spheres model for liquids and colloids with a $24\%$
polydispersity. Extrapolating by \ac{FSS} to the thermodynamic
limit the results obtained from the Maxwell construction in finite
systems, we show that the critical temperature for the
amorphous-crystal phase-separation is {\em below} the dynamic glass
transition, which makes dynamically difficult (although not impossible
\cite{colloids:Zaccarelli09}) to observe such phase-separation. 

At low temperatures the system divides spatially into an amorphous and a
crystalline part, in agreement with previous
findings~\cite{poly:Fernandez07}. The phase-separated amorphous is a stable
fluid {\em below its dynamic glass temperature,\/} which is an optimal
candidate to suffer a thermodynamic glass transition. On the other hand, the
phase-separated solid displays crystalline order. Polydispersities on the
coexisting amorphous and solid are smaller than in the fluid. In fact,
particles distribute spatially according to their size following a complex
pattern not described by any fractionation scenario known to us. However, we
must mention that there are strong similarities with the results of very recent
isobaric semigrand canonical simulations~\cite{sollich:10}.  Although
restricted to smaller system sizes ($N=256$) and polydispersities ($\delta<
7\%$ in the solid phase), these authors find as well that in the crystal phase
the correlations between the fluctuating local particle-sizes extend to quite
long spatial distances.

\acknowledgments We acknowledge BIFI cluster and CINECA for 2$\times
10^5$ hours of computer time. We have been partly supported through
Research Contract Nos. FIS2006-08533, FIS2009-12648-C03-01,
FIS2008-01323 (MICINN, Spain) and by UCM-Banco de Santander (GR58/08).
B.S. was supported by the FPU program (Spain).

\bibliographystyle{apsrev}

\end{document}